      [1999/12/01 v1.4c Il Nuovo Cimento]
\documentclass{cimento}

\usepackage{graphicx}  
\title{The Liverpool Telescope Automatic Pipeline for Real-time GRB Afterglow Detection}
\author{A.~Gomboc\from{ins:x}\from{ins:y}\thanks{andreja.gomboc@fmf.uni-lj.si}, A.~Monfardini\from{ins:x},
 C.~Guidorzi\from{ins:x}, C.~G.~Mundell\from{ins:x}\ETC,, \\
 C.~J.~Mottram\from{ins:x},S.~N.~Fraser\from{ins:x}, R.~J.~Smith\from{ins:x}, 
I.~A.~Steele\from{ins:x}, D.~Carter\from{ins:x}, \\
M.~F.~Bode\from{ins:x} \atque  A.~M.~Newsam\from{ins:x} 
}

\instlist{\inst{ins:x} Astrophysics Research Institute, Liverpool John Moores University, UK
    \inst{ins:y} Faculty of Mathematics and Physics, University in Ljubljana, Slovenia}
  
\PACSes{\PACSit{95.55.Cs}{Ground-based ultraviolet, optical and infrared telescopes}
\PACSit{95.75.Mn}{Image processing (including source extraction)}   
\PACSit{95.75.Rs}{Remote observing techniques}   
\PACSit{98.70.Rz}{gamma-ray sources; gamma-ray bursts}}
\begin{document}

\maketitle

\begin{abstract} 
The 2-m robotic Liverpool Telescope (LT) is ideally
suited to the rapid follow-up of unpredictable and transient events
such as GRBs. Our GRB follow-up strategy is designed to identify optical/IR counterparts 
in real time; it involves
the automatic triggering of initial observations, on receipt of an
alert from Gamma Ray Observatories HETE-2, INTEGRAL and Swift,
followed by automated data reduction, analysis, OT identification and
subsequent observing mode choice. The lack of human intervention in
this process requires robustness at all stages of the procedure. Here
we describe the telescope, its instrumentation and GRB pipeline.
\end{abstract}

\section{GRB follow-up strategy with the Liverpool Telescope}
The LT has a 2-m primary mirror, final focal ratio f/10,
altitude-azimuth design, image quality $<$ 0.4" on axis, 
rapid slew rate of 2$^\circ$/sec and five instrument ports (4 folded and one straight-through, selected by a deployable, 
rotating mirror in the AG Box within 30 sec).
The telescope began science operations in January 2004 and has entered the robotic (unmanned) 
operation phase with an automated scheduler in summer 2004.
At present, instrumentation (Table \ref{tableLT}) includes Optical and Infrared imaging cameras. A prototype low resolution 
spectrograph will be commissioned in 2005 and a higher resolution spectrograph is being developed for 2006.

\begin{table}[htb]
\caption{The Liverpool Telescope instrumentation }
\label{tableLT}
\begin{tabular}{ll}
\hline
{\it RATCam} Optical CCD Camera - & 2048$\times$2048 pixels, 0.135"/pixel, FOV 4.6'$\times$4.6', \\
  & 8 filter selections (u', g', r', i', z', B, V, ND2.0) \\
 & - from LT first light, July 2003 \\
 \hline
 {\it SupIRCam} 1 - 2.5 micron Camera - & 256$\times$256 pixels, 0.4"/pixel, FOV 1.7'$\times$1.7', \\
  (with Imperial College) & Z, J, H, K' filters - from late 2004 \\
  \hline
 {\it Prototype Spectrograph} - & 49$\times$1.7" fibres, 512$\times$512 pixels, R=1000; \\
  (with University of Manchester) & 3500 $<$ $\lambda$ $<$ 7000 \AA - from 2005 \\
  \hline
 {\it FRODOSpec} Integral field  & R=4000, 8000; \\
  double beam spectrograph - & 4000 $<$ $\lambda$ $<$ 9500 \AA - from 2006 \\
  (with University of Southampton) & \\
\hline
\end{tabular}\\[2pt]
\end{table}

As a National Facility, the LT carries out observations for a diverse
range of time-domain related astronomy programmes. However, its robotic
control and automated scheduling make the LT  especially suitable for
rapid response to Targets of Opportunity, such as GRBs (particularly
the short bursts, see \cite{ref:gom} for details) which have been assigned a
high priority in the LT's science programme.

\begin{figure}
\begin{center}
\includegraphics[height=.61\textwidth]{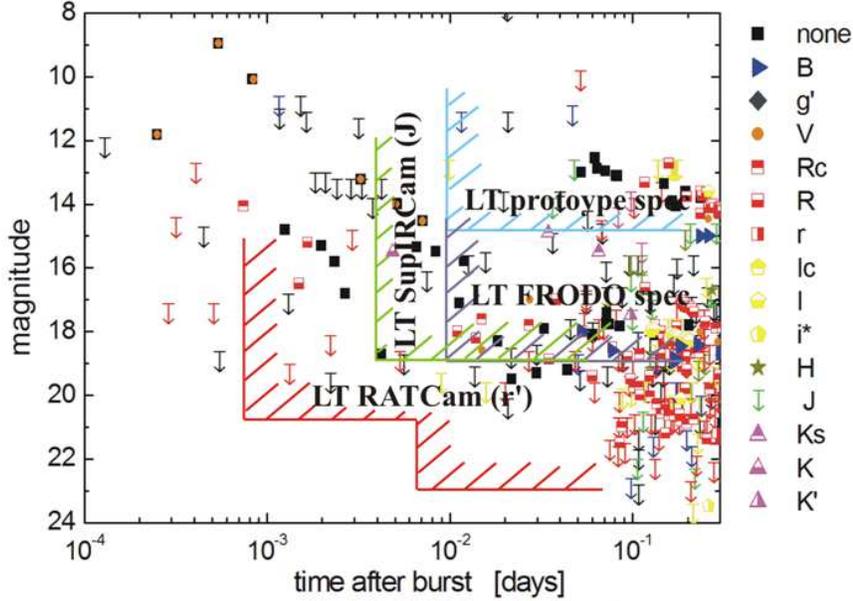}    
\caption{Optical and infra-red observations of GRB afterglows in first minutes and hours after 
the burst (from GCN Circulars): symbols mark detections and arrows upper limits.
Shadowed regions indicate areas where the LT and its instrumentation is expected 
to make a valuable contribution.}
\label{gom2}
\end{center}
\end{figure}

Following a GRB alert from the GRB Coordinates Network, a special
over-ride mode interrupts the current observations and triggers an
initial GRB observing sequence as quickly as 1 minute after receipt of
the satellite alert. As this process is fully automatic, a
crucial component of the system is the subsequent pipelined data
reduction, analysis and automatic identification of possible GRB
counterparts. The diverse range of instrumentation available allows a
number of possible follow-up strategies depending on the burst
characteristics. The automatic choice of follow-up strategy depends on
the output from the pipeline, which is therefore required to be robust
and to reliably detect (or rule out) candidate counterparts in real
time. In the future, it will be possible to include  gamma/X-ray
properties in the automatic choice of strategy; currently however
the deciding factor is the observed optical properties
(e.g. magnitude, variability - see Fig.~\ref{gom2}).
 
The basic overview of the GRB follow-up strategy with the LT 
is as follows:
(i) slew to the position given in GCN alert;
(ii) start short optical exposures in about 1 min (reaching limiting 
magnitudes of r$^\prime \approx$ 20 -- 21 in t$_{\rm{exp}}$= 10 sec -- 1 min);
(iii) automatic pipeline tries to identify the optical transient; 
(iv-a) if no candidate afterglow in the optical detected, continue observations with longer exposures
in optical and infrared (t$_{\rm{exp}}$=10 min giving r$^\prime \approx$ 23 and t$_{\rm{exp}}$=5 min 
limiting J $\approx$ 19);
or (iv-b) if an optical transient is reliably identified:
continue with multicolour optical/infrared imaging and, following the successful commissioning of
the LT spectrographs, with subsequent spectroscopy, provided the optical transient is brighter 
than magnitude 15 for prototype spectrograph and 19 for
FRODO spectrograph (in t$_{\rm{exp}}$= 10 min). 

Alternatively, the follow-up can proceed using the detailed/improved information on the afterglow 
location and magnitude from GCN (Swift, HETE-2, etc). 

\section{GRB pipeline description}

The selection of the observing strategy in the LT robotic telescope is automatically performed 
through the activation of a suitable, preliminary 'detection mode'. The detection mode is a fixed 
procedure designed to avoid saturation even in extreme, GRB 990123-like cases. 
The aim is to fully exploit the capabilities of a large robotic
telescope and understand quickly whether the afterglow is bright. Briefly, three short integration exposures 
(e.g. t$_{exp}$ = ~10 sec.; filter: r$^\prime$) are taken and analysed in real-time to:
\begin{itemize}
\item{extract sources via a customised version of {\it sextractor 2.3.2}~\cite{ref:ber};}
\item{elaborate a list of known sources using a local copy of the USNO-B1.0 astrometric catalog~\cite{ref:mon};}
\item{perform a precise (sub arc-sec) astrometric fit;}
\item{iterate over extracted sources to find possible matches to catalogued ones;}
\item{compile a list of associated sources, and use it to calculate a magnitude offset to be 
applied to instrumental values;}
\item{compile a list of unmatched sources (raw candidates list);}
\item{review the 'raw candidates list' to rate each object according to a number of options 
(e.g. elongation, FWHM, distance to a bright star, extraction smoothness etc.);}
\item{adjust the ratings given taking into account the results of on-line queries to 2MASS, GSC2.3 
and Minor Planet Database;}
\item{cross-check the three 'refined' lists to locate common objects and/or variable objects;}
\item{compile a 'master list' of rated candidates representing the result of the detection mode 
run;}
\item{sort the final, master list and elect the winner, if any;}
\item{output the process details into a human readable file to allow
the astronomer on duty to quickly release a GCN Circular where appropriate.}
\end{itemize}

The whole process takes usually a few seconds to complete. The telescope is kept in idle mode in 
the meantime, waiting for the next scheduled instructions. 
At present, 17 different checks are performed during the 'raw candidates list' to 'master list' 
process. This number is expected to increase in future releases, and the software has been 
designed to allow easy integration of new checks. Among the implemented checks, a powerful one is 
the variability test. Preliminary simulations demonstrate positive results in case of a power-law 
decay ($\alpha$ = -1 -- -2) of a r$^\prime \approx$ 16--17 object observed at (t-t$_{\rm{0}}$) $<$ 200 sec.
Further optimisation is 
foreseen taking into account real afterglow behaviours as soon as they will become available.

In case no object passes the threshold check, the detection is declared failed and the 
'faint' follow-up mode is triggered. The faintest, matched star magnitude (automatically calibrated 
to USNO-B field objects) gives an immediate estimate of the r$^\prime$ afterglow flux upper limit. 
The subsequent schedule makes use of both RATCam and SupIRCam cameras to search for the faint transient 
or to put stringent optical-IR upper limits. 

If one or more objects survives the severe selection, the most probable one is assigned the role 
of 'official candidate'. Depending on its r$^\prime$ magnitude, different strategies are activated making 
use of  the available instruments (spectrograph, SupIRCam, RATCam). Either spectroscopy 
(brightest cases) or multi-colour optical-IR imaging are possible.

\section{Conclusions} 
Rapid reaction to GRB alerts and the automatic
pipeline for afterglow detection combined with 2-m aperture, excellent
site and a range of instrumentation and follow-up strategies enable
the LT to make early and deep observations of GRB optical and infrared
transients (Fig.~\ref{gom2}), and thereby contribute significantly to
the study of early-time light curves, optically-dark GRBs and short
GRBs. Early identification of GRB counterparts with the LT will also be used to
trigger further observations on other facilities.

\acknowledgments
The Liverpool Telescope is funded via EU, PPARC and JMU grants and the generous benefaction of Mr
Aldham Robarts.
AG and CG acknowledge their Marie Curie Fellowships from the European Commission. 
CGM acknowledges the financial support from the Royal Society and MFB is grateful to the 
UK PPARC for provision of a Senior Fellowship.

\end{document}